\documentclass[a4paper, 12pt]{scrartcl}

\usepackage[utf8x]{inputenc}
\usepackage[T1]{fontenc}
\usepackage[english]{babel}

\makeatletter
\DeclareOldFontCommand{\rm}{\normalfont\rmfamily}{\mathrm}
\DeclareOldFontCommand{\sf}{\normalfont\sffamily}{\mathsf}
\DeclareOldFontCommand{\tt}{\normalfont\ttfamily}{\mathtt}
\DeclareOldFontCommand{\bf}{\normalfont\bfseries}{\mathbf}
\DeclareOldFontCommand{\it}{\normalfont\itshape}{\mathit}
\DeclareOldFontCommand{\sl}{\normalfont\slshape}{\@nomath\sl}
\DeclareOldFontCommand{\sc}{\normalfont\scshape}{\@nomath\sc}
\makeatother

\usepackage[a4paper,  centering, bindingoffset=0mm, 
inner=25mm, outer=25mm, top=26mm, bottom=36mm, heightrounded]{geometry}

\usepackage{graphicx}
\usepackage{float}
\usepackage{cite}

\usepackage{amsmath}	
\usepackage{amssymb}
\usepackage{amsfonts}
\usepackage{mathrsfs}
\usepackage{mathtools}
\usepackage{amsthm}

\usepackage{braket}
\usepackage{slashed}
\usepackage{hyperref}

\usepackage{epsf}
\usepackage{epsfig}	
\usepackage{epstopdf}
\usepackage[font=small,labelfont=bf, width=.95\textwidth]{caption}

\usepackage{verbatim}
\usepackage{textcomp}
\usepackage{enumitem}

\usepackage[toc]{appendix}

\usepackage{layout}

\usepackage{microtype}
\usepackage{bookmark}

\usepackage{color}
\usepackage[]{xcolor}

\newcommand{\Ann}{\text{Ann}}
\newcommand{\ii}{\mathrm{i}} 

\newcommand{\sfA}{\mathsf{A}}

\makeatletter
\newcommand{\subalign}[1]{%
  \vcenter{%
    \Let@ \restore@math@cr \default@tag
    \baselineskip\fontdimen10 \scriptfont\tw@
    \advance\baselineskip\fontdimen12 \scriptfont\tw@
    \lineskip\thr@@\fontdimen8 \scriptfont\thr@@
    \lineskiplimit\lineskip
    \ialign{\hfil$\m@th\scriptstyle##$&$\m@th\scriptstyle{}##$\crcr
      #1\crcr
    }%
  }
}
\makeatother

\definecolor{unamblue}{cmyk}{1 0.79 0.12 0.59}

\usepackage[]{hyperref}
\hypersetup{
    colorlinks=true,%
    citecolor=unamblue,%
    filecolor=unamblue,%
    linkcolor=unamblue,%
    urlcolor=unamblue,
    bookmarksnumbered=true,     
    bookmarksopen=true,         
    bookmarksopenlevel=1,       
    pdfstartview=Fit,           
    pdfpagemode=UseOutlines,
    pdfpagelayout=TwoPageRight
}

\usepackage{graphicx}
\usepackage{tikz}
\usepackage{slashed}
\usepackage{epstopdf}
\usepackage{verbatim}	
\usepackage{blkarray}
\usepackage{tikz}
\usepackage{ytableau}
\usepackage{tikz-feynman}
\tikzfeynmanset{warn luatex=false}

\usepackage[	]{todonotes}

\usepackage{subcaption}

\title{\Huge Holonomic representation of biadjoint scalar amplitudes}

\author{\\[5mm] \normalfont\normalsize Leonardo de la Cruz\\[2mm]
	\emph{\normalfont\normalsize \em Institut de Physique Th\'eorique,  CEA, CNRS, Universit\'e Paris–Saclay,}\\
	\emph{\normalfont\normalsize \em  F-91191, Gif-sur-Yvette cedex, France 	}
}
\date{%
	$\,$%
	\\[2\baselineskip]
	\normalfont\normalsize%
		\parbox{0.8\linewidth}{%
				{\bf \sf Abstract}. 
		We study  tree-level biadjoint scalar amplitudes in the language of $D$-modules. We construct left ideals  in the Weyl algebra $D$ that allow a holonomic representation of $n$-point amplitudes  in terms of the linear partial differential equations they satisfy. The resulting representation 
		 encodes the simple pole  and  recursive properties of the amplitude.  		}
}

\begin{document}
\maketitle
\thispagestyle{empty}

\newpage

\section{Introduction}
 Given a system of $q$ linear partial  differential equations for a function $f(x)=f(x_1, \dots , x_N)$ 
 \begin{equation}
 P_i(x_1, \dots ,x_N, \partial_{x_1},\dots, \partial_{x_N} ) f(x)=0,\quad i=1,\dots, q \,
 \end{equation}
the study of the system and  $f$ itself can be translated into the study of their \emph{annihilating} operators 
$P_1, \dots, P_q$. The variables $x_1, \dots, x_N$, the differential operators $\partial_{x_1}, \dots, \partial_{x_N}$, the commutations rules $[x_i, x_j]=[\partial_{x_i}, \partial_{x_j}]=0$, $[\partial_{x_i}, x_j]=\delta_{ij}$  generate a non-commutative ring known as the \emph{Weyl algebra} $D_N$, usually denoted simply by $D$. Thus, 
 the operators $P_i$ are  generators of a (left) ideal $I$ in   $D_N$ and 
 the system of differential equations can be seen as  a module over  $D_N$(a $D$-module).
  When the dimension of this module  is minimal\cite{zbMATH03369365}, namely $N$, the function $f$  is said to be 
holonomic. Among other properties, holonomic functions can be constructed from other  holonomic functions  by performing operations such 
as addition, multiplication, and integration.
A related concept is the so called  \emph{canonical holonomic representation} of $f$, introduced by Zeilberger \cite{ZEILBERGER1990321}, which  consists of a holonomic ideal
together with a certain set of initial conditions that determine $f$ uniquely. 
   The relation between $D$-modules and  holonomic functions has  been comprehensible reviewed in  Ref.\cite{2019arXiv191001395S}. On the computational side, $D$-module methods\footnote{The main concepts of the theory can be consulted in math textbooks Refs.\cite{sturmfels:1999,coutinho_1995} or in Ref.\cite{Weinzierl:2022eaz} for a more physicist-oriented literature.}  have been implemented for various Computer Algebra Systems (CAS) incorporating the algorithms described in Refs.\cite{sturmfels:1999,Koutschan09}, in particular the construction of annihilator ideals.

The theory of $D$-modules has  been applied to the study of differential equations of Feynman integrals and their relation with hypergeometric functions \cite{Kalmykov:2020cqz}.  $D$-modules are at the core of the identification of   
Feynman integrals as $\sfA$-hypergeometric functions 
 \cite{delaCruz:2019skx, Klausen:2019hrg}, a perspective that has been further developed in  Refs.\cite{Klausen:2019hrg, Tellander:2021xdz, Klausen:2021yrt, Ananthanarayan:2022ntm, Klausen:2023gui,Chestnov:2022alh, Dlapa:2023cvx, Chestnov:2023kww} (see also Refs.\cite{Zhang:2023fil, Klemm:2019dbm,Bonisch:2020qmm, Bonisch:2021yfw, Mizera:2021icv, Lairez:2022zkj, Henn:2023tbo}).  The  ideals  associated with Feynman integrals are holonomic and thus Feynman integrals are holonomic functions  \cite{Kashiwara:1977nf}. Because holonomicity is preserved under integration, it can be used to derive  relations between  Feynman integrals in parametric representation  and establish other interesting 
 properties \cite{Bitoun:2017nre} (see also Refs. \cite{Fujimoto:2012vfa,Vanhove:2018mto}). The point of view of $D$-modules has also appeared in the  construction of  annihilators for  amplitudes in Yang-Mills and gravity \cite{Nutzi:2019ufl} motivated by conformal symmetries in tree-level graviton  amplitudes \cite{Loebbert:2018xce}. 

In this paper, we will apply basic aspects of the $D$-modules to  tree-level biadjoint scalar amplitudes. Amplitudes are rational functions so they are  holonomic functions. This means that we  can construct their  canonical holonomic representations by  deriving holonomic left ideals (the annihilator $D$-ideal) and set boundary conditions that determine them completely.

The remainder of this paper is organized as follows. In Section \ref{preliminary} we review the recursive definition of biadjoint scalar amplitudes and basic notions of $D$-modules. In Section \ref{D-mods-biadjoint} we compute their annihilators and construct their holonomic representations. Our conclusions are presented in Section \ref{conclusion}.
\section{Review}
\label{preliminary}
\subsection{Notation}
 Amplitudes for $n$ massless particles depend on kinematic (Mandelstam)   invariants  defined by
\begin{align}
	s_{ijk\dots}:=(p_i+p_j+p_k+\dots)^2, 
\end{align} 
where the outgoing massless momenta are subject to momentum conservation $\sum_{i=1}^n p_i^\mu=0
$ and on-shell conditions $p_i^2=0$. 
Momentum conservation and on-shell conditions
set the number of  independent variables as
\begin{equation}
	N:= \frac 12 n(n-3) \, .
\end{equation}
We will omit the momentum dependence in double-ordered scalar amplitudes and write $m_n(w| \tilde w):=m_n(p,w, \tilde w)$. For $\tilde w=w$ we define $m_n(w):=m_n(w|w)$ and for the standard ordering $w=123\dots n$ we omit the ordering and simply write $m_n$.  
 We utilize the notation $m_n(S)$ to emphasize that we view the amplitude as a  function of   $N$ Mandelstam invariants. We write $\partial_{x_i}:=\partial/\partial x_i$ and for Euler operators 
 $\theta_{x_i}:=  x_i\partial_{x_i}$.
\subsection{Biadjoint scalar amplitudes}
The basic description of  biadjoint scalar amplitudes starts with a lagrangian density
\begin{align}
	\mathcal L=  \frac{1}{2} \partial_\mu\varphi_{a \alpha} \partial^\mu \varphi_{a \alpha}
	-\frac{\lambda}{3!} f^{abc}
	\tilde f^{\alpha\beta\gamma}
	\varphi_{a\alpha} \varphi_{b\beta}\varphi_{c\gamma}\, ,
	\label{lagrangian} 
\end{align}
where the structure constants $f^{abc}$ and $\tilde f^{\alpha\beta\gamma}$ are associated to the gauge groups $U(\mathrm{N})$ and 
$U(\tilde {\mathrm{N}})$, respectively\footnote{Generators are normalized according to $[T^a, T^b]=\ii f^{abc} T^c$ and 
	$[\tilde T^{\alpha}, \tilde T^{\beta}]=\ii \tilde f^{\alpha \beta \gamma} \tilde T^{\gamma}$.}.  Full $n-$point amplitudes have a double-color decomposition into  traces depending on the generators and  tree-level double-ordered primitive (partial) amplitudes $m_n(\sigma| \tilde \sigma)$
given by
\begin{align}
	\mathcal M_n(p) = \lambda^{n-2}  \sum_{\sigma\in S_n/ \mathbb Z_n}
\sum_{\tilde \sigma
	\in S_n/ \mathbb Z_n}
	\text{Tr} (T^{a_{\sigma(1)}} \cdots T^{a_{\sigma(n)}}) \text{Tr}(T^{a_{\tilde \sigma(1)}} \cdots T^{a_{\tilde \sigma(n)}}) m_n(\sigma| \tilde \sigma) \,, 
\end{align}
where $\sigma$ and $\tilde \sigma$ denote cyclic orderings. We will focus on  the partial amplitudes $m_n(\sigma| \tilde \sigma)$ for the remainder of this work.  
There are various equivalent ways of representing them,  including the Cachazo-He-Yuan representation\cite{Cachazo:2013hca,Cachazo:2013iea}, canonical forms \cite{Arkani-Hamed:2017mur}, and  intersection numbers \cite{Mizera:2017rqa,delaCruz:2017zqr}. 
However, their basic
 definition is in terms of Feynman diagrams.
Let $G$ be a diagram. We denote by $E(G)$ the set of internal edges and by $s_e$ the Lorentz invariant corresponding to the internal edge $e$.  Partial amplitudes $m_n(\sigma| \tilde \sigma)$
 are then given by 
\begin{equation}
	m_n(\sigma| \tilde \sigma)=(-1)^{n-3+n_{\text{flip}}(\sigma,\tilde \sigma)} \sum\limits_{G\in \mathcal T_n (\sigma) \cap\mathcal T_n (\tilde \sigma)}\prod\limits_{e\in E(G)}\frac 1 {s_{e}}\, ,
	\label{Feyn-graph}
\end{equation}
where $\mathcal T_n (\sigma) \cap\mathcal T_n (\tilde \sigma)$ denotes the set of all trivalent graphs compatible with the external orderings $\sigma$ and $\tilde \sigma$, and $n_{\text{flip}}(\sigma,\tilde \sigma)$ is
the number of flips needed to transform any diagram from $\mathcal T_n (\sigma) \cap\mathcal T_n (\tilde \sigma)$ with the external ordering $\sigma$ into another with external ordering $\tilde \sigma$, see Refs. \cite{Cachazo:2013iea, delaCruz:2016wbr} for more details. 

A more convenient definition for our purposes  is through the Berends-Giele recursion \cite{Berends:1987me} proposed in 
Ref.\cite{Mafra:2016ltu}.
Let us briefly review it  in order to introduce some notation.
External cyclic orderings can be identified with ordered sequences of letters (words) in an alphabet $\mathbb A_n=\set{1, \dots, n}$. Words of length $n$ are sequences of letters  $l_i\in \mathbb A_n$ of the form  $w=l_1 \cdots l_n$. The empty word is denoted by $e$ and  the Mandelstam invariant of a word $w$ of length 
$|w|$ is defined by $s_w:=s_{{l_1} {l_2}\cdots l_{|w|}}$.
Let $w$ be a word in the alphabet $\mathbb A_n$
 and let $\sum_{xy=w}$ be the sum over all possible ways to deconcatenate the word $w$ into two non-empty words $x$ and $y$. The recursion for the  amplitude is constructed from 
 \begin{align}
 	\phi_{w_1,w_2}=\frac{1}{s_{w_1}}
 	\sum_{xy=w_1}
 	 	\sum_{ab=w_2}
 	 	\left[
 	 	\phi_{x,a}\phi_{y,b}- (x\leftrightarrow y)
 	 	\right], \quad \phi_{w_1,w_2}\equiv0, \quad \text{if} \quad w_1 \setminus w_2 \ne  e\,  
 	 	\label{rec-1}
 	\end{align}
 with the start of the recursion defined as $\phi_{i,j}=\delta_{ij}$. The $n$-point amplitude is
\begin{align}
	m_n(w_1n| w_2n)=(-1)^{(n-3)}s_{w_1} \varphi_{w_1,w_2} \, ,
\label{rec-2}
	\end{align}
where the double cyclic invariance has been used to set the orderings to $w = w_1 n$ and
$\tilde w = w_2 n$.
Choosing the canonical ordering 
$w=123 \dots n $, amplitudes up to $n=5$ are given by 
\begin{align}
	m_3=1, \quad	m_4=-\frac{1}{s_{12}}-\frac{1}{s_{23}}, \quad
	m_5= \frac{1}{s_{12} s_{123}}+\frac{1}{s_{12} s_{34}}+\frac{1}{s_{123} s_{23}}+\frac{1}{s_{23} s_{234}}+\frac{1}{s_{234} s_{34}} \, .
	\label{ampsupto5}
\end{align}
Higher-point amplitudes can be obtained recursively from Eqs.\eqref{rec-1}-\eqref{rec-2}. 
A  {\sf Mathematica}  implementation of this recursion  is given in Appendix \ref{code-biadjoints}.

The functional form of the amplitude is  independent of the labels so we will use Eq.\eqref{rec-2} to define  (sub)-amplitudes when letters $l_i$ are replaced by sub-words $w_i$. Let $w=w_1 w_2\dots w_{n-1} w_{n}$, then the sub-amplitude for $w$ is given by the evaluation of Eq.\eqref{rec-2} with the replacements $l_i \to w_i, 1\le i\le n-1$ and $w_n=e$.
 For instance, let us take $w=w_1 w_2 w_3 w_4$, 
   then the corresponding 4-point sub-amplitude reads
\begin{align}
	m_4(w_1,w_2,w_3 |w_1,w_2,w_3 )=-\frac {1}{ s_{w_1 w_2}}-
	\frac {1} {s_{w_2 w_3}} \, ,\label{four-point-relabel}
\end{align} 
where we have set $w_4=e$.  The sub-amplitudes we are  considering also appear in the algorithm of Ref.\cite{Cachazo:2013iea} to compute biadjoint scalar amplitudes by drawing polygons.

\subsection{D-modules and holonomic functions}
We will keep this section short and refer the reader to 
Chapter 6 of \cite{hibi2014groebner} and Ref.\cite{2019arXiv191001395S} for details.  
The $N$-th Weyl algebra with complex coefficients is the ring of differential operators $\partial_{x_1},\dots, \partial_{x_N}$ with coefficients in the polynomial ring $\mathbb C[x_1, \dots, x_N]$
\begin{align}
	D_N:=\mathbb C [x_1,\dots, x_N]\braket{\partial_{x_1}, \dots,\partial_{x_N}},
\end{align}
which is a non-commutative algebra  generated by $x_1, \dots, x_N$, $\partial_1, \dots, \partial_N$ modulo the commutations relations $[x_i, x_j]=[\partial_{x_i}, \partial_{x_j}]=0$ and $[\partial_{x_i}, x_j]=\delta_{ij}$. Using the commutation relations, any element of $D_N$ can be expressed uniquely in a basis of normal ordered \emph{monomials}
\begin{align}
x^\alpha \partial^\beta=	x_1^{\alpha_1} \cdots x_N^{\alpha_N} \partial_{x_1}^{\beta_1} \cdots \partial_{x_N}^{\beta_N} \, ,
\label{ordered-monomials}
	\end{align}
where the differential operators appear at the rightmost position. Here $\alpha,\beta \in \mathbb N^N$ are exponent vectors, where the length is given by $|\alpha|=\alpha_1+\dots+\alpha_N$ and similarly for $\beta$. 
 We will also consider left ideals where the coefficients of  the operators are rational expressions in $N$ variables. The latter can be expressed in general as $f/g$, $g\ne 0$, where $f,g$ are polynomials in $N$ variables with complex coefficients. The ring of differential operators with rational function coefficients is defined by
\begin{align}
	R_N:=\mathbb C(x_1, \dots, x_N) \braket{\partial_{x_1}, \dots,\partial_{x_N}},
	\end{align}  
where $\mathbb C(x_1, \dots, x_N)$ is the field of rational expressions in $N$ variables. In addition to the commutation rules for $D_N$, in $R_N$ the multiplication of the operator $\partial_{x_i}$ and $a(x)\in \mathbb C(x)$ is defined by
\begin{align}
	\partial_{x_i} a(x):=a(x)\partial_{x_i} +
	\partial_{x_i} a(x)\, .
	\end{align}
Any element of $R_N$ can be expressed in terms of a basis of the form $ a^\alpha(x) \partial^\beta$, where as in Eq.\eqref{ordered-monomials} the operators $\partial_i$ are at the rightmost position.
Any element of $R_N$ acts on a function as 
\begin{align}
	a(x) \partial_{\alpha} \bullet f(x)= a(x) 
	\frac{\partial^{|\alpha|}f(x)}{\partial x_1^{\alpha_1} \cdots\partial x_N^{\alpha_N}}\, ,
\end{align}
where the symbol $\bullet$ is used to distinguish it from the multiplication in $R_N$. For $p, q \in R_N$, we have 
\begin{align}
	(pq)\bullet f=p\bullet(q\bullet f)\, . \label{property}
\end{align}

Now we will give some definitions and a proposition whose proof can be found
in  Ref.\cite{ZEILBERGER1990321} and Chapter 20 in \cite{coutinho_1995}.

\subsubsection*{Definitions}
\begin{description}
	\item \emph{Holonomic function}.  A left ideal is called holonomic if its (Bernstein) dimension is the smallest possible, namely $N$.
Let $f$ be a function and  consider all elements in $D_N$ that annihilate $f$
 \begin{align}
 \text{Ann}_{D_N}(f)=\set{P \in D_N: P\bullet f=0} \,.
 \label{ideal-general} 
  \end{align} 
A function  $f$ is called holonomic if its annihilator  $\text{Ann}_{D_N}(f)$ is a holonomic ideal.
\item \emph{Canonical holonomic representation (Zeilberger)}. A canonical holonomic representation is given by the ideal \eqref{ideal-general} together with the following set of initial conditions. Let $\alpha_1, \dots, \alpha_n$ denote the orders of the operators in $ \text{Ann}_{D_N}(f)$, then 
set $\alpha_1 \cdots \alpha_N$ initial conditions
\begin{align}
	\partial_{x_1}^{i_1} \cdots
		\partial_{x_N}^{i_N} \bullet f|_{x=x_0}, \quad 0\le i_k < \alpha_k, \quad \text{for} \quad k=1, \dots, N, \label{boundaries}
\end{align}
	where $x_0$ is any point that is not in the characteristic set of the system, namely the set of common zeros of the leading coefficients of the operators $P_i$. 
\end{description}
\subsubsection*{Proposition}
\begin{description}
\item  \emph{Let $f$ and $g$ be holonomic functions in $N$ variables, then $1/f$, $1/g$, $f g$, and $f+g$ are holonomic functions.\footnote{Other  operations that preserve holonomicity are  convolution, restriction and both indefinite and definite integration \cite{ZEILBERGER1990321}.}} 
\end{description}
\subsubsection*{Example}
Consider the function $f(x,y)=y e^{-x^2}$ whose annihilator ideal is given by
\begin{align}
	I=\braket{2x+\partial_x, y\partial_y-1}. 
\end{align}
The system of PDEs associated with it is denoted by $L_1 \bullet f(x,y)=L_2 \bullet f(x,y)=0$, where $L_1$, $L_2$, are the generators of the left ideal $I$. The generators of the ideal are of order $1$ and hence we need to set a single boundary condition, which we can choose as $f(0,1)=1$.

\section{Holonomic biadjoint scalar amplitudes}
\label{D-mods-biadjoint} 
A corollary of the the above proposition is that
\emph{biadjoint scalar  amplitudes are holonomic functions}. To prove it  notice that each contribution of the amplitude, say the Feynman-diagram based representation \eqref{Feyn-graph} is a sum of inverse products of Mandelstam invariants of the form $s_{12}s_{23} \dots $, which are themselves holonomic and so are their inverses and sums. This is of course true for any rational function.

Holonomicity of biadjoint amplitudes implies that we can find a representation of them as left ideals, or in other words as solutions of a system of partial differential equations. Moreover, imposing a set of initial conditions on the amplitude $m_n(S)$, thought as a function of $N$ variables, we can construct a canonical holonomic representation of $m_n$ as defined above.
In order to construct such representation we need to establish first the Weyl algebra corresponding to $n$-point amplitudes. For this purpose let us define the ring on which polynomial coefficients of operators live. 
Focusing on $m_n(123\dots | 123\dots)$, we notice that the recursive structure of the amplitude based on Eq.\eqref{rec-2} induces a basis $B_n$ of words at each $n$ given by
\begin{align}
	B_n =\set{ w \in \text{Part}_2 \ \mathbb A_{n-1}} \cup 
	\set{w \in \text{Part}_3 \ \mathbb A_{n-1}}\cup \dots \cup
	\set{w \in \text{Part}_{n-2} \ \mathbb A_{n-1}} \, ,
	\label{basis}
	\end{align} 
where  $\text{Part}_{i} \mathbb A_k$ is the overlapping partition of a $\mathbb A_k$ with offset 1 and length $i$. We have the chain of inclusions
\begin{align}
	B_3\subseteq B_{4}	\subseteq \dots 	\subseteq B_{n-1} 	\subseteq B_{n}\, ,
\end{align}
where by definition $B_3$ is the empty set. Therefore, kinematic invariants will be labeled by $S_n=\set{s_{w}|w\in B_n}$, where $|S_n|=\frac 12 n(n-3)=N$, so its associated ring is
$\mathbb C[S_n]$. Other bases can be equally acceptable as long as they contain $N$ elements, see e.g., Ref.\cite{Cheung:2017pzi}.
 A more formal derivation of this ring can be found in Ref.\cite{Henning:2017fpj}.
We then define the corresponding   set of operators by 
$\partial_{S_n}:=\set{\partial_{s_w} | w\in B_n}$ so the  associated Weyl algebra 
 is 
\begin{align}
	D_N=\mathbb C[S_n]\braket{\partial_{S_n}} \,.
\end{align}
We are now interested in constructing the annihilator ideal $I_n:=\Ann_{D_N} (m_n)$ of $m_n$. 
For a polynomial $f^\alpha(x)$ we can always construct annihilators of the form 
\begin{align}
	f\partial_i -\alpha (\partial_i f)\in 	\Ann_{D_N}(f^\alpha), \quad \text{for} \quad  
	i=1, \dots, N\, .
	\label{annis-simple}
\end{align} 
Scattering amplitudes can be expressed as rational functions of the form $m_n=f/g$ so by analogy with \eqref{annis-simple} we can construct simple  annihilators of the form
\begin{align}
	Q_i= m_n  \partial_i - \frac 1 g 
	\left[-m_n \partial_i g  + \partial_i f\right], \quad  i=1, \dots, N,
	\label{rational}
\end{align}
where $Q_i\in R_N $. 
Equivalently, using Eq.\eqref{property}, we have
\begin{align}
	 P_i=g f \partial_i +(f\partial_i g-g\partial_i f), \quad  i=1, \dots, N, \label{easy-annis}
\end{align}
where $P_i\in D_N$.
Moreover, from the identity $\theta_x\bullet (1/x)=-1/x$ it is  easy to deduce that $m_n$ is annihilated by
\begin{align}
H_n=\left[\sum_{w\in B_n } \theta_{s_w}  + (n-3)\right]\, , 
\label{anniP}
\end{align}
which is a consequence of the simple pole structure of amplitudes. Therefore, a possible left ideal of  $(N+1)$
generators is
\begin{align}
 \braket{P_1, \dots, P_N, H_n}  \subset D_N\, .
\label{ideal-naive}
\end{align}  
The holonomicity of $m_n$ implies that we can do better and construct a left ideal with exactly $N$ generators (see Section 4 of Ref.\cite{ZEILBERGER1990321}). We may obtain one by, say, dropping $H_n$ or a single $P_i$. We can of course  start with the annihilators \eqref{rational} and construct an ideal in $R_N$ instead.
A more systematic way of obtaining these ideals is through CAS. The task of computing annihilators in general difficult  but several implementations already exist, e.g, the $\sf{Mathematica}$ package 
 $\sf{HolonomicFunctions}$ \cite{Koutschan09,Koutschan10b,Koutschan10c} or  the {\sf Macaulay2} package "Dmodules".
\subsection{4-points}
At $4$-points Eq.\eqref{basis} gives $B_4= \set{w\in \text{Part}_2 \mathbb A_3}= \set{12, 23}$ and therefore
$S_4=\set{s_{12}, s_{23}}$. The associated Weyl algebra is then $D_2=\mathbb C [s_{12}, s_{23}] \braket{\partial_{s_{12}},\partial_{s_{23}}  }
:=\mathbb C[S_4] \braket{\partial_{S_4}}$.
The left ideals that annihilate the amplitude are not unique so it is interesting to compare them.
Let us consider e.g. Eq.\eqref{ideal-naive} for
 $n=4$ and dropping
$H_4$. The corresponding ideal is
\begin{align}
	\text{Ann}_{D_2}(m_4)=\braket{(s_{12}+s_{23})\theta_{s_{12}}+s_{23},(s_{12}+s_{23})\theta_{s_{23}}+s_{12}}\,, 
	\label{ann-4-pt-bare}
	\end{align}
which is also the output of $\sf{HolonomicFunctions}$. Let us compare it against the representation computed from {\sf Macaulay2 }
\begin{align}
	\text{Ann}_{D_2}(m_4)=
	\braket{\partial_{s_{12}} \partial_{s_{23}}, s_{12}\partial_{s_{12}} +s_{23}\partial_{s_{23}}+1, s_{23}\partial_{s_{23}}^2+2\partial_{s_{23}}
	} \, ,
\label{m2-annihilator}
	\end{align}
which  has three generators and where the maximum order is two. These representations can be shown to be equivalent after performing a left Gr\"obner basis computation of Eq.\eqref{m2-annihilator}\footnote{ The computation of Gr\"obner bases for differential operators is outside the scope of this work.  The  {\sf HolonomicFunctions} package has the command {\sf OreGroebner} for this purpose. In the examples we use DegreeLexicographic order. }. 
The same is true for   ideal composed by three generators $\braket{P_1,P_2,H_4}
$, which can also be reduced to 
 Eq.\eqref{ann-4-pt-bare}. 

Now, acting on the left with $1/s_{12}$ and $1/s_{23}$ on  generators of the ideal \eqref{ann-4-pt-bare}, respectively, we can  rewrite  it as
\begin{align}
I_4=&\braket{
		m_4s_{12}\theta_{s_{12}}-1,  	m_4s_{23}\theta_{s_{23}}-1},
	\label{anns-4pt}
\end{align}
where strictly speaking the annihilator 
now belongs to $R_{2}$.
 Despite the fact that the annihilator depends on the function we wish to represent, the system of differential equations have the simple form
\begin{equation}
	\begin{aligned}
		s_{12}\theta_{s_{12}} m_4(S)=1,	\\
		s_{23}\theta_{s_{23}} m_4(S)=1.
	\end{aligned}
\end{equation}
The amplitude is not determined uniquely by this pair of differential equations since we have not imposed boundary conditions. Since the order of the 
generators is one  we need to set 
a single boundary condition (see Eq.\eqref{boundaries}) which we choose as
 $\lim_{s_{12},s_{23}\to\infty} m_4(s_{12},s_{23})=0$. The ideal \eqref{anns-4pt} and the boundary condition constitute a canonical holonomic representation of  $m_4$.

Before going to higher points, 
let us briefly consider annihilators of other partial amplitudes, say, $m_4(1234|1243)$ and 
$m_4(1234|1423)$. We have  
\begin{align}
	m_4(1234|1243)=\frac{1}{s_{12}} \Rightarrow
	\Ann_D(m_4(1234|1243))=\braket{\theta_{s_{12}}+1},\\
	m_4(1234|1423)= \frac{1}{s_{23}}
	\Rightarrow
	\Ann_D(m_4(1234|1423))=\braket{\theta_{s_{23}}+1},
\end{align}
which we can express as
\begin{align}
	\Ann_D(m_4(1234	|1243))=&\braket{m_4(1234|1243)s_{12}\theta_{s_{12}}+1},\\
	\Ann_D(m_4(1234|1423))=&\braket{m_4(1234|1423)s_{23}\theta_{s_{23}}+1},
\end{align}
and the differential equations read
\begin{align}
	s_{12}\theta_{s_{12}} m_4(1234|1243)=-1,\\
	s_{23}\theta_{s_{23}} m_4(1234|1423)=-1,
\end{align}
respectively. 
Amplitudes with orderings different from $\tilde w=w=1234$ only contain subset of poles and thus this property is also reflected on the annihilators. In general, it is possible to construct the annihilator and the holonomic representation of, say, $f+g$ if the representations of $f$ and $g$ are known. Instead, here we will directly focus on the annihilators of  $m_n(12 \dots n|12 \dots n)$. 

\subsection{5-points}
At $5$-points the basis  is given by
\begin{align}
 B_5= \set{12,23,34}\cup \set{w\in \text{Part}_3 \mathbb A_3}= \set{12, 23, 34,123, 234}
 \end{align}
  so
the associated Weyl algebra is
 $D_5=\mathbb C (S_5)\braket{\partial_{S_5}}$. Starting with $I_5=\braket{P_1, \dots, P_5}$ the  annihilator  ideal  can be brought into the form  
\begin{align}
  	I_5=
  	 & \Big\langle  \frac{1}{s_{123}}+\frac{1}{s_{34}}+m_5 s_{12} \theta _{s_{12}},\frac{1}{s_{123}}+\frac{1}{s_{234}}+m_5 s_{23} \theta _{s_{23}}, \frac{1}{s_{12}}+\frac{1}{s_{234}}+m_5 s_{34} \theta _{s_{34}},  \\
  	& \quad\frac{1}{s_{12}}+\frac{1}{s_{23}}+m_5 s_{123} \theta _{s_{123}}, \frac{1}{s_{23}}+\frac{1}{s_{34}}+m_5 s_{234} \theta _{s_{234}}
\Big\rangle \, .
\nonumber
\end{align}
 This form is equivalent to the ideal obtained from ${\sf HolonomicFunctions}$ and also ${\sf Macaulay2}$ after a Gr\"obner bases computation.
The rational terms appearing in the annihilators have the functional form of a 4-point amplitude. Indeed, the rational terms of the  generators of $I_5$ are given by
\begin{equation}
\begin{aligned}
	\frac{1}{s_{123}}+\frac{1}{s_{34}}=&-m_4(12,3,4),\quad \frac{1}{s_{12}}+\frac{1}{s_{23}}=-m_4(1,2,3), \quad  
	\frac{1}{s_{123}}+\frac{1}{s_{234}}=-m_4(1,23,4),\\
	\frac{1}{s_{23}}+\frac{1}{s_{34}}=&-m_4(2,3,4), \quad \frac{1}{s_{12}}+\frac{1}{s_{234}}=
	-m_4(1,2,34),
\end{aligned} 
\end{equation}
where we have used Eq.\eqref{four-point-relabel}. Hence, defining
\begin{align}
	\sfA_5 :=&  m_5 \ \text{diag} 
	\left( s_{12}, \ s_{23} ,  \ s_{34}, \ s_{123} , \ s_{234}   \right), \\
	\theta_5:=& (\theta_{s_{12}},\theta_{s_{23}}, \theta_{s_{34}},\theta_{s_{123}},\theta_{s_{234}})^{T},\\
 \qquad \kappa_5:=& (m_4(12,3,4), m_4(1,23,4), m_4(1,2,34), m_4(1,2,3), m_4(2,3,4))^{T}\ ,
\end{align} 
we can express $I_5$  as $\braket{\sfA_5  \theta_5- \kappa_5}$, which together with the initial condition $m_5|_{S_5\to \infty}=0$ give a holonomic representation of $m_5$.

\subsection{Higher points}
We follow the same procedure at higher points. Starting with $I_n=\braket{P_1, \dots, P_N}$ we divide over a factor of Mandelstam invariants, which can be read off from the product $s_w g$, for $w\in B_n$ (see Eq.\eqref{rational}).   We find that the annihilator ideal with $N$ generators of the   $n$-point amplitude is
\begin{align}
	I_n=\braket{\sfA_n  \theta_n-\kappa_n}, \label{general-ideal}
\end{align}
where the $N\times N$  diagonal matrix $\sfA$ and the vector of parameters are given by
\begin{align}
	\sfA_n= m_n\text{diag}(s_{12}, s_{23} , \dots ), \quad \theta_n=(\theta_{s_{12}},
	\theta_{s_{23}}, \dots  )^T, \quad \kappa_n= (\iota_n(12), \iota_n(23) , \dots)^T, 
	\label{general-ideal-A}
\end{align}
respectively. The ellipsis indicates all remaining words in the basis $B_n$ and the function
$\iota(w)$ is defined by
 \begin{align}
	\iota_n(w):= m_{|w+1|}(l_1\dots l_{|w|}) m_{n-|w|+1}(1,2, \dots, w, \dots, n-1)
\label{iotafunct}
\end{align}
for $2\le |w| \le n-2$ and zero otherwise. Since all  operators in $I_n$ are of order one we  need a single boundary condition as indicated by 
Eq.\eqref{boundaries}, which we choose as
$m_n|_{S_n\to \infty}=0$.  This ideal and the boundary condition then specifies a canonical holonomic representation of the $n$-point amplitude. For $n>5$,  we find  
nonlinear entries in $\kappa_n$, which depend of  lower point sub-amplitudes as can be seen from Eq.\eqref{iotafunct} (see Appendix \ref{examples-kappa} for examples). The ideal in Eq.\eqref{general-ideal-A} implies that the amplitude $m_n	$ thought as a function of $N$ variables satisfies the system of differential equations
\begin{align}
s_w \theta_{s_w} m_n(S)=\kappa_w, \quad \forall w\in B_n  \, .
\label{diff-eqs}
\end{align}
We have checked  Eqs.\eqref{general-ideal}-\eqref{diff-eqs} up to $n=10$.

\section{Conclusions}
\label{conclusion}
We have studied double-ordered biadjoint amplitudes in the language of $D$-modules. Their  holonomicity implies that there exist a canonical representation of $n$-point amplitudes made up of an annihilator left ideal with precisely $N$ generators, which we have constructed explicitly.  We have found that in general the annihilator can be constructed from a diagonal matrix which depends on the independent kinematic invariants and a vector made up of lower point sub-amplitudes.

It would be interesting to study canonical holonomic 
representations of   generalizations of biadjoint amplitudes, which have been proposed in Refs.\cite{Cachazo:2022pnx,Cachazo:2022vuo}.  Similar Berends-Giele recursions to those used here also exist for Yang-Mills 
\cite{Mafra:2015vca} and were also important for the derivation of second order differential operators in Ref.\cite{Nutzi:2019ufl} so it would be interesting to construct canonical holonomic representations for these amplitudes and determine whether the  resulting differential equations manifest a dependence on sub-amplitudes.%
\addsec{Acknowledgements}
We thank Carlos Mafra for helpful comments on the  manuscript. 
This work is  supported by 
the European Research Council under grant ERC-AdG-885414.

\appendix
\section{Mathematica code to calculate biadjoint scalar amplitudes}
\label{code-biadjoints}
\begin{verbatim}
	SetAttributes[s,Orderless]
	xAB[A__,B__]:=If[Complement[A,B]=={},xAB[A,B],0];
	xAB[{i_Integer},{j_Integer}]:=If[i==j, 1, 0];
	xAB[A__,B__]:=xAB[A,B]=(1/s[A])Plus@@Flatten[Table[xAB[A[[1;;j]],B[[1;;k]]]
	xAB[A[[j+1;;Length[A]]],B[[k+1;;Length[B]]]]-xAB[A[[j+1;;Length[A]]],B[[1;;k]]]
	xAB[A[[1;;j]],B[[k+1;;Length[B]]]],{j,1, Length[A]-1},{k,1, Length[B]-1}]];
	
(*Standard ordering 123...n-1 *)
mAB[A__, B__] := (-1)^(Length[A] - 2) xAB[A, B] (s @@ A) /. {List -> 
	Sequence} // Expand;
	\end{verbatim}

\section{Explicit values of $\kappa_n$}
\label{examples-kappa}
Here we give some explicit values of $\kappa_n$ up to $n=7$. From Eq.\eqref{basis} we have
\begin{align}
B_6=&\{12,23,34,45,123,234,345,1234,2345\}, \\
	B_7=&\{12,23,34,45,56,123,234,345,456,1234,2345,3456,12345,23456\}.	 
\end{align}
Hence, using Eq.\eqref{iotafunct} the corresponding values of 
	$\kappa$ are
\begin{align}
\kappa_6=&\\
&[m_5(12,3,4,5),m_5(1,23,4,5),m_5(1,2,34,5),m_5(1,2,3,45),m_4(1,2,3) m_4(123,4,5),\nonumber\\
&m_4(2,3,4) m_4(1,234,5),m_4(3,4,5) m_4(1,2,345),m_5(1,2,3,4),m_5(2,3,4,5)],\nonumber\\
	\kappa_7=&\\
	&[m_6(12,3,4,5,6),m_6(1,23,4,5,6),m_6(1,2,34,5,6),m_6(1,2,3,45,6),m_6(1,2,3,4,56),\nonumber \\
	&m_4(1,2,3) m_5(123,4,5,6),m_4(2,3,4) m_5(1,234,5,6),m_4(3,4,5) m_5(1,2,345,6),\nonumber\\
	&m_4(4,5,6) m_5(1,2,3,456),m_5(1,2,3,4) m_4(1234,5,6),m_5(2,3,4,5) m_4(1,2345,6),\nonumber\\
	&m_5(3,4,5,6) m_4(1,2,3456),m_6(1,2,3,4,5),m_6(2,3,4,5,6)],\nonumber
	\end{align}
respectively, where we  have used $m_3(w_1,w_2)=1$. 

\bibliographystyle{JHEP.bst}

 \renewcommand\bibname{References} 
\ifdefined\phantomsection		
  \phantomsection  
\else
\fi
\addcontentsline{toc}{section}{References}

\providecommand{\href}[2]{#2}\begingroup\raggedright\endgroup


\end{document}